\documentstyle[prl,aps]{revtex}
\begin{document}       
\twocolumn[{
\draft
\widetext

\title
{Low-energy relativistic effects and nonlocality in time-dependent
tunneling }

\author{Gast\'on Garc\'{\i}a-Calder\'on}
 
\address{Instituto de F\'{\i}sica,
Universidad Nacional Aut\'onoma de M\'exico\\
Apartado Postal 20-364, 01000 M\'exico, Distrito Federal, M\'exico}

\author{Alberto Rubio\cite{ar}}
 
\address{Facultad de Ciencias,
Universidad Aut\'onoma de Baja California\\
Apartado Postal 1880, 22800 Ensenada, Baja California, M\'exico}

\author{Jorge Villavicencio\cite{jv}}

\address{Centro de Investigaci\'on Cient\'{\i}fica y de Educaci\'on
Superior de Ensenada\\
Apartado Postal 2732, 22800 Ensenada, Baja California, M\'exico}

\date{19 August 1998}

\mediumtext

\begin{abstract}
We consider exact time-dependent analytic solutions to the
Schr\"odinger equation for tunneling in one dimension with
cut off wave initial conditions at $t=0$.
We obtain that as soon as $t \neq 0$ the transmitted probability
density at any arbitrary distance rises instantaneously with time
in a linear manner.
Using a simple model we find that the above  nonlocal effect of
the time-dependent solution is suppressed by consideration of
low-energy relativistic effects. Hence at a distance $x_0$
from the potential the probability density rises after a time
$t_0=x_0/c$ restoring Einstein causality. This implies that
the tunneling time of a particle can never be zero.
\end{abstract}
\maketitle
}]

\narrowtext
Recent technological achievements as the possibility of
constructing artificial quantum structures at nanometric
scales\cite{qs} or manipulating individual atoms\cite{corral}
have stimulated a great deal of work at both an applied and
fundamental level.
In particular, studies on tunneling have addressed, among
others, the controversial
question of the traversal time of a particle through a classically
forbidden region\cite{traversal}. The above considerations
have motivated
a renewed attention to the time-dependent treatments of quantum
tunneling. From the theoretical side, most of these works are
based on the numerical analysis of the time-dependent
Schr\"odinger equation with the initial condition of
a Gaussian wave packet\cite{gaussian}.
A common feature in most of these approaches is that the
initial wave packet extends through all space. As a consequence
the initial state, although it is manipulated to reduce as much as
possible its value along the tunneling and transmitted regions,
contaminates from the beginning the tunneling process.
In the literature, however, one also finds a number
of approaches to time-dependent tunneling, pionnered by
Stevens\cite{stevens83},
that in fact circumvent the above situation  using a cut off wave as
initial state\cite{stevens83,morettipra92,muga96,gcr97}.

Our approach is a generalization to an arbitrary
potential\cite{gcr97} of the Moshinsky shutter\cite{mm}.
Moshinsky considered the solution
of the time-dependent free Schr\"odinger equation with the initial
condition, at $t=0$, of a plane wave of momentum $k$ confined
in the half-space region $x < 0$ to the left of a perfectly
absorbing shutter located at $x=0$.
The sudden opening of the shutter at time $t=0$, allows the
plane wave solution to propagate freely along the region $x > 0$
\cite{note1}.
Moshinsky showed that as the time $t$ goes to infinity, the
solution to the problem tends to the stationary solution. He also found
that both the current and the probability density for a fixed value
of the distance $x_0$ as a function of $t$, present oscillations
near the wavefront, situated at $t_0=x_0/v$. He named this phenomenon
diffraction in time, in analogy to the well known phenomenon of
optical diffraction.
Recently an observation of diffraction in time has been
reported\cite{exp}. If we put a potential barrier in the
region $0 \leq x \leq L$ with the same initial condition as above,
then we may have a convenient model to analyze tunneling times by
measuring at what time the probability density rises from zero.
However, as pointed out by Holland\cite{holland} for the free
case, and by Garc\'{\i}a-Calder\'on and Rubio\cite{gcr97},
for the case of a potential, the solution
of the time-dependent Schr\"odinger equation for a cut off initial
plane wave has a nonlocal character.
This means that if initially there is a zero probability for the
particle to be at $x > 0$, as soon as  $t \neq  0$, there is
instantaneously a finite, though very small, probability to find it
at any point $x > 0$.  This implies a zero tunneling time for some
particles.

In this work we address the issue of the behaviour of the
time-dependent solution to the Schr\"odinger equation for tunneling
through a potential barrier using a cut off wave as initial condition.
Our aim is to analyze the nonlocal behaviour of the
time-dependent transmitted solution at early times.
We also study low-energy relativistic
effects  by solving the Klein-Gordon equation for a model potential.
The implication of our findings for the tunneling time problem is
briefly discussed.

For the sake of simplicity in our approach we consider
the instantaneous removal of the shutter. This may be seen
as a kind of `sudden approximation' to a shutter opening with
finite velocity, where the treatment is more involved
even for the free case\cite{gahler}.
In a recent paper we have shown that the transmitted solution for the
Schr\"odinger case for tunneling through an arbitrary potential barrier
may be written as a free term solution plus a infinite sum
resonance transient terms associated with the S-matrix poles of the
problem\cite{gcr97}. This corresponds to solve the time-dependent
Schr\"odinger equation for a potential $V(x)$ that vanishes outside
the region $0 \leq x \leq L$, with the initial condition,
\begin{equation}      
\psi_s(x,k,t=0)=\left\{ \begin{array}{cc} e^{ikx},
& x<0 \\ 0, & x>0.
\end{array}      
\right.
\label {2}
\end{equation}
The transmitted solution for the region $x \geq L$ reads\cite{gcr97},
\begin{equation}
\psi_s(x,k,t) = T(k)M(x,k,t) -i \sum_n^{\infty} T_n M(x,k_n,t),
\label{3}
\end{equation}
where $T(k)$ stands for the transmission amplitude of the problem,
$ T_n= u_n(0)u_n(L){\rm exp}(-ik_nL)/(k-k_n)$, is given in terms of
the resonant eigenfunctions $u_n(x)$ and complex S-matrix poles $k_n$;
and the Moshinsky functions $M(x,k,t)$ and $M(x,k_n,t)$ are defined as,
\begin{equation}
M(x,q,t)= \frac{1}{2}{\rm e}^{(imx^2/2\hbar t)}
{\rm e}^{y^2}{\rm erfc}(y),
\label{4}
\end{equation}
where the argument $y$ is given by
\begin{equation}
y \equiv {\rm e}^{-i\pi/4}\left ( \frac {m}{2\hbar t} \right )^{1/2}
\left [x- \frac {\hbar q}{m}t \right ].
\label{5}
\end{equation}
In the above two equations $q$ stands either for $k$ or $k_n$.
In the absence of a potential the solution given by Eq. \ (\ref{3})
becomes the solution for the free case obtained by Moshinsky\cite{mm},
\begin{equation}
\psi_s^0(x,k,t)= M(x,k,t).
\label{5a}
\end{equation}
As discussed by Moshinsky, the initial condition given
by Eq.\ (\ref{2}) refers to a shutter that acts as a perfect
absorber (no reflected wave). One can also envisage a shutter that
acts as a perfect reflector. In such a case the
initial wave may be written as,
\begin{equation}
\psi_s(x,k,t=0)=\left\{ \begin{array}{cc} e^{ikx}-e^{-ikx},
& x<0 \\ 0, & x>0.
\end{array}      
\right.
\label {2a}
\end{equation}
The transmitted solution for the region $x \geq L$ now reads,
\begin{eqnarray}
\psi_s(x,k,t) =&& T(k)M(x,k,t)-T(-k)M(x,-k,t) \nonumber\\
&&  -2ik \sum_n^{\infty} T_n M(x,k_n,t),
\label{3a}
\end{eqnarray}
where $T_n=u_n(0)u_n(L){\rm exp}(-ik_nL)/(k^2-k_n^2)$.
The solution for the free case with the reflecting initial condition is
\begin{equation}
\psi_s^0(x,k,t)= M(x,k,t)-M(x,-k,t).
\label{5b}
\end{equation}
The exact solutions given by Eqs.\ (\ref{3}) and (\ref{3a}),
corresponding, respectively, to absorbing and reflecting initial
cut off waves, involve each a contribution proportional to the free
case
solution and then an infinite sum involving the S-matrix poles, $\{k_n\}$,
and resonant states, $\{u_n(x)\}$, of the system.
As shown in ref.\ \cite{gcr97}, at very long times, the terms
$M(x,k_n,t)$ that appear in the above equations vanish. The same occurs
for $M(x,-k,t)$ while, as shown firstly in ref. \cite{mm}, $M(x,k,t)$
tends to the stationary solution. Hence, at long times, each of the above
exact solutions go into the stationary solution
$T(k){\rm exp}[i(kx-Et/\hbar)]$.

At very short times, for a given $x \geq L$, the argument
of $M(x,k,t)$, given by Eq.\ (\ref{5}) with $q=k$, becomes very large
and in fact becomes independent of the value of $k$,
$y \approx {\rm exp}(-i\pi/4)[m/2\hbar t]^{1/2}x$.
Since for very large $y$, $M(y) \sim 1/y$ \cite{mm,gcr97}, it follows
that $M(x,k,t)$ goes like $t^{1/2}$. As discussed also in ref.
\cite{gcr97}, the functions dependent on the poles, $M(x,k_n,t)$,
behave in a similar fashion provided the value of $t=t_0$ is
sufficiently small to guarantee, for a fixed $x=L$, that
$L \gg \hbar|k_n|t/m$.
Since the distribution of the complex S-matrix poles on the
k-plane fulfills $[|k_1| < |k_{2}| ... < |k_n| ...,$ one sees
that as $t$ becomes smaller and smaller there will be
more and more values of $n$ for which the corresponding $M$ functions
goe like $t^{1/2}$  as do all the rest of $M$ functions associated with
smaller values of $n$. In the appropriate limit
as $t \rightarrow 0$ and $n \rightarrow \infty$,
the corresponding $M$ function then vanishes as $t^{1/2}$.
Consequently for $x \geq L$, the solutions given by Eqs.\ (\ref{3})
and (\ref{3a}) are proportional to $t^{1/2}$, namely,
\begin{equation}
\psi_s(x,k,t) \sim \frac{A}{x} t^{1/2},\,\,\,\,\, (x \geq L)
\label{5c}
\end{equation}
where $A$ a constant. Note that at $t=0$ the solution vanishes
in accordance with the initial condition.
It is not difficult to see that Eq.\ ({\ref{5c}) will hold also
for a cut off initial condition that is something between the initial
conditions considered above, and more generally, for a wave packet
formed by a linear combination of cut off waves.
Eq.\ (\ref{5c}) tell us that the probability density
at any distance $x$ from the potential will rise instantaneously
with time.
This intriguing nonlocal behaviour implies that an ideal detector
will measure a zero tunneling time.
The existence of action-at-a-distance effects in the time-dependent
Schr\"odinger equation should not in principle pose any conceptual
difficulties since the treatment is non-relativistic.
However one could ask whether the above nonlocal behaviour
arises because the initial condition is a cut off wave.
In order to answer the above question we consider low-energy
relativistic effects by solving the Klein-Gordon equation
with a cut off wave as initial condition for a simple potential
model. Moshinsky\cite{mm} solved the Klein-Gordon equation
for the free shutter problem with the initial condition of a cut off
plane wave in the region $x < 0$ and showed that the probability
density at a point $x > 0$ is nonzero only after a time $t_0 > x_0/c$,
with $c$ the velocity of light.
To our knowledge a numerical analysis of this solution has not yet
been performed. We would like to learn also how the relativistic
solution is affected at early times by tunneling through a potential.

A potential that has been widely used in studies on time-dependent
tunneling is the square barrier, characterized by a height $V_0$
and a width $L$. This potential has an infinite set of S-matrix poles
situated at increasing energies on top of the barrier.
There is, however, a simpler potential model that
is more amenable for a relativistic treatment.
This is the delta potential $V(x)=b_s\delta(x)$.
The solution corresponding to the time-dependent Schr\"odinger
equation has been obtained by Elberfeld and Kleber using
a delta-function propagator\cite{ek}.
One can also follow a derivation by Laplace transforming
directly the time-dependent Schr\"odinger equation of the problem
using the initial condition given by Eq.\ (\ref{2})\cite{gcmm}.
Defining $p^2=2ims/\hbar$ the Laplace transformed solution
$\overline{\psi}_s(x,s)$ for the region $x > 0$ reads,
\begin{equation}      
\overline{\psi}_s(x,k,s)= \frac {im}{\hbar}\frac {e^{i px}}
{\left(p+ i b \right) \left( p-k\right) },
\label{6}
\end{equation}
where $b=mb_s/\hbar^2$.
After a simple partial fractions decomposition the inverse Laplace
transform yields for $x>0$,
\begin{equation}
\psi _s^\delta (x,k,t)=T(k)M(x,k,t)+R(k)M(x,-ib;t).
\label{7}
\end{equation}
where T(k) and R(k) stand for the transmission and reflection
amplitudes for the stationary situation,
$T(k)=k/(k+ib)$ and $R(k)=ib/(k+ib)$. Note that here instead
of an infinite number of S-matrix poles the only S-matrix pole
corresponds to an antibound state located at $k_a=-ib$.
At a very short times one can easily see that $\psi_s^\delta(x,k,t)$
goes like $t^{1/2}$ fulfilling also, as the square barrier,
Eq.\ (\ref{5c}).

The shutter problem for the Klein-Gordon equation
with the delta potential $V(x)=b_r \delta (x)$
requires the solution of
\begin{eqnarray}
\frac{\partial ^2}{\partial x^2}\psi_r^{\delta} (x,k_r,t)
=&&\frac {1}{c^2}\frac{\partial^2}{\partial t^2} \psi_r^{\delta}
(x,k_r,t) \nonumber\\
+&& [b_r \delta (x)+\mu ^2] \psi_r^{\delta}(x,k_r,t)
\label{15}
\end{eqnarray}
where $\mu =mc/\hbar$.
 with the initial condition given by,
\begin{equation}          
\psi_r^{\delta} (x,k_r,t=0)=\left\{
\begin{array}{cc}
e^{ik_rx}, & x<0 \\
0, & x>0
\end{array}        
\right.
\label{9}
\end{equation}
where we define $E_r^2=k_r^2+\mu ^2$ and $k_r=k(1-(k/\mu)^2)^{-1/2}$.
Note that $E_r$ is given in units of the reciprocal length, i.e.,
$E_r \equiv E/\hbar c$.
The condition given by Eq.\ (\ref{9}) follows from the fact
that for $t < 0$, when the shutter was closed, we had on the left
side of the shutter, $\psi^{\delta}(x,k_r,t)={\rm exp}[i(k_rx-E_rct)]$,
for $x <0$ and a vanishing value for $x > 0$.
By direct application of the Laplace transform
method one gets a set of differential equations corresponding
to the regions $x>0$ and $ x<0$. In order to derive an expression
for the transmitted wave function, we
have to consider the matching conditions to take into account the
discontinuity of the wave function derivatives at $x=0$, obtaining
the Laplace-transformed solution,
\begin{equation}      
\overline{\psi }_r(x,s)=\frac 12\frac{(s-icE_r)}{\left(
q+b_0 c \right) \left( q+ik_rc \right) }e^{-qx/c}.
\label{16}
\end{equation} 
where $q=[s^2+\mu^{2}c^{2}]^{1/2}$ and $b_0 =b_r/2$.
Using the Bromwich contour
to evaluate the inverse Laplace transform of Eq.\ (\ref{16})
it is convenient to make the change of variable\cite{mm}
$-iu=(q+s)/(\mu c)$. In this form $q=i\mu c(u^{-1}-u)/2$ and
as a consequence the branch points at $s=\pm i\mu c$ go into
an essential singularity at $u=0$ and two simple poles
located on the lower half of the complex $u$-plane. After separating
into partial fractions one then may evaluate the resulting integrals
by standard complex variable techniques to obtain the wave function,
\begin{equation}
\psi_r^{\delta} (x,k_r,t)=\left\{
\begin{array}{cc}
& A\psi _r^0(x,k_r,t) + B C \psi _r^0(x,-ib_0,t)+ \nonumber\\
& BD[\psi _r^{0}(x,- ib_0 ,t)]^*,\,\,\, t>x/c\\
& 0,\,\,\,\, t<x/c
\end{array}
\right.
\label{17}
\end{equation}
where $A=k_r/(k_r+ib_0)$, $B=ib_0/(k_r+ib_0)$,
$C=(\epsilon+E_r)/(2\epsilon )$, and $D=(\epsilon -E_r)/(2\epsilon)$,
and also $\epsilon =(\mu ^2-b_0^2)^{1/2}$.
In Eq. \ (\ref{17}), the function $\psi_r^0(x,k_r,t)$ is the
solution of the free Klein-Gordon case\cite{mm,dirac}, namely,
\begin{equation}
\psi_r^0 (x,k_r,t)=\left\{
\begin{array}{cc}
&e^{i(k_rx-E_rct)}+\frac 12J_0(\eta )- \nonumber\\
&\sum\limits_{n=0}^\infty \left[\xi /iz\right] ^nJ_n(\eta ),\,\,\,\,
t>x/c\\
&0,\,\,\,\, t<x/c
\end{array}
\right.
\label{12}
\end{equation}
where $J_n(\eta )$ stands for the Bessel function of order $n$ and,
\begin{equation}      
\xi =\left[ \frac{ct+x}{ct-x}\right] ^{1/2}, \eta =\mu
(c^2t^2-x^2)^{1/2}\,, z=\frac 1\mu (k_r+E_r).
\label{13}
\end{equation}
The expressions, $\psi_r^0(x,-ib_0 ,t)$ and
$(\psi _r^0(x,-ib_0 ,t))^*$ in Eq.\ (\ref{17})
have the same form as the free solution in Eq.\ (\ref{12})
with $k_r$ replaced by $-ib_0$. Asymptotically for
very long times in the solution $\psi_r^{\delta} (x,k_r,t)$,
given by Eq.\ (\ref{17}), the terms $\psi _r^0(x,-ib_0 ,t)$ and
$(\psi _r^0(x,-ib_0 ,t))^*$  vanish, while the
term $\psi _r^0(x,k_r ,t)$ goes into the stationary solution
${\rm exp}[i(k_rx-E_rct)]$.

To exemplify the above results Fig. \ref{fig1} exhibits the very short
time behaviour of the probability density for the delta potential
at a fixed distance $x=L=0.3 \AA$ . One sees that the Schr\"odinger
description (broken line), obtained from Eq.\ (\ref{7}) with parameters
$b_s=2.0$ ${\rm e}V-\AA$ and $E=0.01$ ${\rm e}V$,
yields an instantaneous response with time while the relativistic
solution, calculated using Eq.\ (\ref{17}), starts after $t_0=L/c$.
This tell us something relevant: The nonlocal behaviour of the
Schr\"odinger description is due to its non-relativistic nature.
The nonlocal behavior of the Schr\"odinger solution would result
from the fact that in a non-relativistic description there is no
restriction on the velocity of some components of the initially
confined wave function. The sharp relativistic wavefront of height
$0.25$ in Fig. \ref{fig1} follows as a consequence of the initial
condition given by Eq.\ (\ref{9}). This jump occurs also in the free
case and may be obtained analytically\cite{mm}. For an initial
function of the type
${\rm exp}(ik_rx)+{\rm exp}(i\alpha){\rm exp}(-ik_rx)$, $(x<0)$,
with $\alpha$ an arbitrary phase, the peak height
will be a function of $\alpha$. In particular for
a reflecting initial condition, $(\alpha=\pi)$, the solution
starts smoothly from zero at $t_0=L/c$.
It might be of interest to mention that in fully relativistic
quantum field theories Hegerfeldt\cite{hegerfeldt}
has pointed out that the sudden opening of a shutter
may lead to violation of Einstein causality, i.e., no propagation
faster than light.
This author has argued that the difficulty is of
a theoretical nature and has discussed some ways to solve it.
Our relativistic model satisfies Einstein causality.
The inset to Fig. \ref{fig1} shows that at longer times the above
two solutions approach each other, both presenting the
characteristic transient behaviour near the `classical' wavefront
at $x=vt$, which in our example occurs at a very short time.
Our analysis  has a consequence of interest for the tunneling time
problem. Since the probability density rises with time
after a time $t_0=x_0/c$, it implies that the tunneling time
of a particle can never be zero, contrary to some claims in the
literature\cite{traversal}.

Thus we can see that a proper description of the quantum mechanical
propagation for the transmitted solution, even at low energies,
strictly requires a relativistic treatment.
However, since the corresponding solutions are practically identical
up to the relativistic cut off, at $t=L/c$, suggests that
the Schr\"odinger description is quite accurate provided the velocity
components larger than $c$ are omitted.

G.G-C. thanks M. Moshinsky for
useful discussions and acknowledges partial
financial supports of SEP-ConacyT, M\'exico, under grant
940085-R97 and DGAPA-UNAM, under grant IN106496.

\begin{figure}     
\caption{ Plot of $|\psi _s^\delta(x,t)|^2$
(dashed line) and $|\psi_r^\delta (x,t) |^2$ (solid line),
respectively, for the Schr\"odinger and Klein-Gordon solutions
for a delta potential, as a function of time at a fixed distance
at early times and at long times (inset). See text.}
\label{fig1}
\end{figure}


\begin{references}
\bibitem{note1} Note that the shutter refers to the initial
condition and hence it is not part of the system.
\end{references}
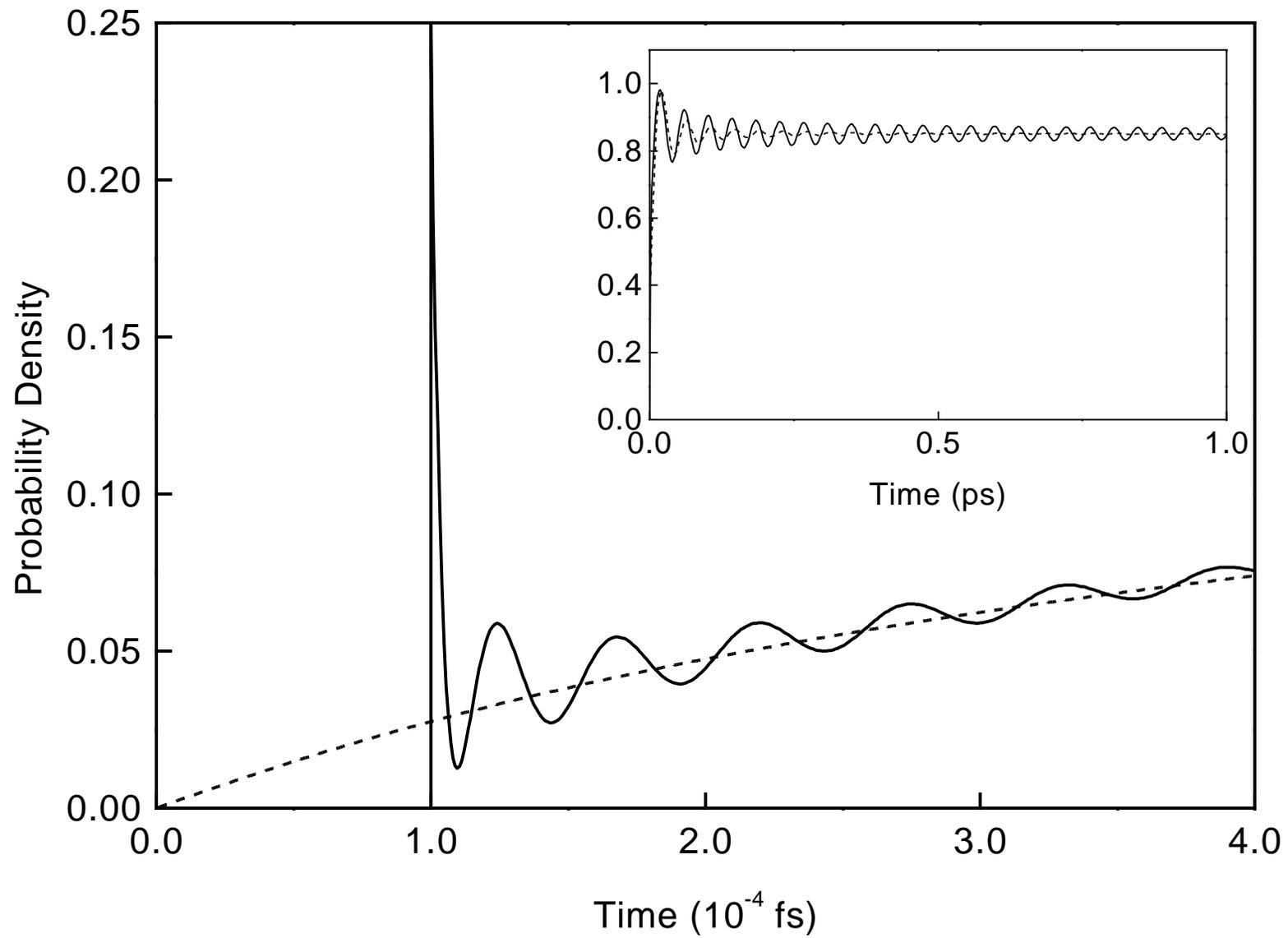
\end{document}